\documentclass[twocolumn,showkeys]{revtex4-1}

\usepackage{graphics}
\usepackage{graphicx}
\usepackage{amsmath}
\usepackage{epsfig}

\begin{document}

\title{Polymer Field-Theory Simulations on Graphics Processing Units}

\author{Kris~T.~Delaney}
\email{kdelaney@mrl.ucsb.edu}
\affiliation{Materials Research Laboratory, University of California, Santa Barbara, CA 93106-5121, USA}
\author{Glenn~H.~Fredrickson}
\email{ghf@mrl.ucsb.edu}
\affiliation{Departments of Materials and Chemical Engineering, University of California, Santa Barbara, CA 93106-5050, USA}

\date{\today}

\begin{abstract}
We report the first CUDA\textsuperscript{\texttrademark} 
graphics-processing-unit (GPU) 
implementation of the polymer field-theoretic simulation
framework for determining fully fluctuating expectation values of equilibrium properties for
periodic and select aperiodic polymer systems.
Our implementation is suitable both for self-consistent field theory (mean-field)
solutions of the field equations, and for fully fluctuating simulations using the
complex Langevin approach.
Running on NVIDIA\textsuperscript{\textregistered} Tesla T20 series GPUs, we 
find double-precision speedups of up to $30\times$ compared to single-core serial calculations on a recent 
reference CPU, while single-precision calculations proceed up to $60\times$ faster than those on
the single CPU core. Due to intensive communications overhead, an MPI implementation running
on $64$ CPU cores remains two times slower than a single GPU.
\end{abstract}


\maketitle

\section{Introduction}
Computational physics has long been a consumer of advanced computational resources and a driver for the development
of cutting-edge hardware.
Rapid strides in development continue today, and recent installations of publicly hosted research supercomputers have begun
breaking the petaFLOP barrier.
However, since single-core CPUs hit the power wall over the last decade, exploiting modern resources
in the physical sciences, which tend to deliver problems that are almost uniquely both bandwidth and compute 
limited, requires increasing levels of software sophistication with multiple layers of carefully designed parallelism.

A recent flurry of activity has taken place in the use of graphics processing
units (GPUs) for general-purpose computing. 
These devices are inherently massively parallel, with modern GPUs containing hundreds of light-weight
cores. 
Single GPUs have recently passed the teraFLOP barrier in total throughput for single-precision
arithmetic.
The field that began with manually packaging non-graphical computations into the language of graphical operations
has matured with the release of general programming frameworks such as NVIDIA's CUDA and OpenCL.
The development of these general programming frameworks, combined with the recent inclusion 
of hardware double-precision operations, has made targeting GPUs an increasingly attractive endeavor
for computational scientists, both as low-heat, low-expense desktop-size cluster replacements, and as 
high-performance co-processors in distributed cluster architectures.
This activity is reflected in the recent uptick of publications reporting implementations and performance
metrics of GPU simulation codes in fields as diverse as electronic structure theory\cite{Esler2010},
computational fluid dynamics\cite{cclw:gpu_cfd_unstructured:2009,Goeddeke:2009:GAO}, molecular dynamics\cite{Anderson2008}, and electromagnetic simulations\cite{Balevic:2008:ASL:1380847.1381216,Klockner:2009:NDG:1613335.1613429}.

Polymer statistical field theory has proven to be a powerful theoretical construct for both
analytical and numerical computations on a wide range of equilibrium polymer systems\cite{GHF}.
Field theories of homogeneous systems have been combined with powerful analytical techniques, such as 
the renormalization group, to elucidate important and fundamental phenomena such as the excluded volume 
effect\cite{FreedBook}. 
Models of inhomogeneous systems, such as phase separated polymer alloys and block copolymers, 
have largely succumbed to  numerical self-consistent field theory (SCFT), which is a computer simulation methodology 
based on a mean-field approximation to the underlying field theory model\cite{SchmidReview1998, MatsenReview2002}.
At the frontier of the field are ``field-theoretic simulations'' (FTS), which attempt to stochastically sample the 
unapproximated (and complex valued) field theory, thereby enabling the study of fluctuation phenomena in both 
homogeneous and inhomogeneous polymer systems\cite{GHF,VenkatReview2002,Ganesan2001}.

In this article, we present details of a GPU implementation of the polymer field theory
framework, including beyond-mean-field simulations, and report favorable runtimes compared 
to the same code running both in serial and in parallel on CPUs.
We have taken care to report benchmark timings together with precise models of the 
hardware involved.
There are limited reports of GPU implementations of other polymer field theory 
simulation methods in the literature. 
Gao and coworkers\cite{Gao2011} recently reported a self-consistent field theory implementation specifically
for semi-flexible block copolymers in two dimensions, with an observed $5.3\times$ 
speedup over their CPU code. 
Wright and Wickham demonstrated a peak $30\times$ speedup on GPUs for a simple
Landau-Brazovskii type free-energy model of diffusive polymer dynamics\cite{Wright2010}.
The analytic and real-valued free-energy model used in the latter avoids the expensive propagator calculations that
are the hallmark of SCFT and the more involved FTS simulations. 
To our knowledge, there has been no prior report in the literature on a GPU implementation of a field-theoretic polymer
simulation.

We note a recent critical review\cite{Lee2010} of reported GPU code speedups by Lee \emph{et al.}
While reports of GPU codes running high-throughput tasks $100\times$ to $1000\times$ faster than their CPU counterparts
are not uncommon, Lee \emph{et al.} attribute such favorable timings to comparisons between carefully optimized
GPU code and less well-optimized CPU code.
For a wide range of commonly used core algorithms, they report comparisons of carefully optimized GPU and CPU code,
with the latter exploiting all of the avenues for on-chip vectorization, effective use of large on-chip caches, and 
appropriate reorganizations of memory access patterns.
They find that CPU code can typically match the runtimes of GPU code to within an order of magnitude, if particular
attention is spent optimizing both codes.
Thus, we have taken care to ensure that our CPU code, which shares much of the code base
with the GPU implementation, is aggressively optimized.
In particular, we do exploit on-chip vectorization in our CPU code, through the use of 
hand-coded streaming SIMD extensions 3 (SSE3) instructions for arithmetic on 
contiguous vectors of complex numbers, and through shared-memory parallelism with OpenMP for exploitation of multi-core architectures. 
We demonstrate in Sec.~\ref{sec:results} that with such careful optimizations,
our runtimes on both CPU and GPU 
are easily dominated by the Fourier transformation steps that are handled 
by high-performance fast Fourier transform (FFT) libraries. 
This dominance is not entirely surprising, since multidimensional FFTs are not trivially parallelizable and are the only component of our calculation that does not scale
linearly with system size.

\section{Theoretical Approach}
A comprehensive discussion of the theoretical foundations of our approach
can be found elsewhere\cite{GHF}.
Here we summarize the most important expressions that underpin our numerical scheme.

We begin by specifying a particle-based model, with the relevant degrees of freedom
corresponding to positions of statistical segments of the polymer chains. 
Interactions between statistical segments are characterized by short-ranged 
intra-molecular interactions (e.g., Gaussian stretching, backbone stiffness, \ldots), and
short- or long-ranged intermolecular interactions (e.g., excluded-volume interaction, 
Flory-like contact potentials, electrostatic interactions, \ldots).
We limit the remainder of our discussion, though not necessarily our simulation code,
to fully flexible Gaussian chains that interact only through contact potentials.
In order to decouple interactions between statistical segments, we introduce
auxiliary fluctuating fields through a Hubbard-Stratonovich transformation and 
integrate out the then independent particle degrees of freedom. 
The resulting canonical partition function is typically of the form
\begin{equation}
\mathcal{Z}_C = \mathcal{Z}_0 \prod_j \int \mathcal{D}\omega_j \mathrm{exp}\left(-H\left[\left\{\omega\right\}\right]\right),
\label{eq:canpartfn}
\end{equation}
where $\mathcal{Z}_0$ is some field-independent constant, $\omega_j\left(\vec{x}\right)$ are the set of
auxiliary ``chemical potential'' fields introduced to decouple interactions, 
$\int \mathcal{D} \omega_i$ denotes a functional integral over such a field, and 
$H$ is a complex-valued energy functional that contains model-dependent interactions.
Other statistical ensembles are readily derived.

The energy functional, $H$, usually contains both explicit and implicit dependences 
on the fields. For example, the Edwards model of homopolymers in an implicit solvent\cite{Edwards1965}, 
which has a single chemical potential field, may be written as
\begin{equation}
H\left[\omega\right] = \frac{B}{2}\int d\vec{x} \left[\omega\left(\vec{x}\right)\right]^2 - C\tilde{V}\ln Q\left[i\omega\right],
\label{eq:ham_edwards}
\end{equation}
where $B$ is a measure of the excluded volume interaction, $C$ is the number of polymer chains per unit volume, $\tilde{V}$
is the system volume in units of the free-polymer radius of gyration, $R_g$, and $Q$ is the normalized partition function 
discussed below.
Similarly, the well-studied case of an incompressible
melt of diblock copolymer consisting of connected blocks of 
incompatible ``A'' and ``B'' segments 
can be written in terms of ``pressure'' and ``exchange'' fields as\cite{MatsenSchick1994,GHF}
\begin{equation}
H\left[\omega_+,\omega_-\right] = C \int d\vec{x}\left[\frac{\omega_-^2}{\chi_{AB}N} - i\omega_+\right] - C\tilde{V} \ln Q\left[\omega_A, \omega_B\right],
\label{eq:ham_db}
\end{equation}
where $\chi_{AB}$ is the Flory contact interaction between statistical segments of ``A'' and ``B'', and $N$ is the total
number of statistical segments in the copolymer chain.
The chemical potential fields $\omega_A$ and $\omega_B$ are the fields felt by ``A'' and ``B'' segments 
respectively, and they are related to the pressure and exchange fields ($\omega_+$ and $\omega_-$) through
the mappings $\omega_A = i\omega_+ - \omega_-$, $\omega_B = i\omega_+ + \omega_-$.

Both of these model Hamiltonian functionals contain implicit non-linear and non-local dependencies 
on the chemical potential fields through the normalized partition function, $Q$, which is derived
from the statistical ensemble of a single polymer chain interacting with the specified chemical
potential field(s). 
All details of polymer architecture are embedded within the $Q$ functional. 
While the present discussion is limited to fully flexible and monodisperse 
linear homopolymer or diblock copolymer chains, the 
method, and indeed our simulation code, is easily extended to more complex architectures such
as symmetric and fully asymmetric multiblock copolymers, branched and star polymer chains, and grafted
copolymers, by simply substituting the appropriate form of $Q$ and introducing fields to decouple any extra
interactions.

The intensive part of our simulation method is the determination of the normalized partition
function, $Q\left[\left\{\omega\right\}\right]$, for a fully specified set of chemical potential
fields.
We write $Q$ in terms of a Feynman-Kac-like propagator, $q$, as
\begin{equation}
Q\left[\omega\right] = \frac{1}{\tilde{V}}\int d\vec{x}\, q\left(\vec{x},s=1;\left[\left\{\omega\right\}\right]\right),
\label{eq:Q}
\end{equation}
where $s$ is the polymer backbone contour variable, which has been normalized by $N$, the
length of the polymer, so that the chain contour length is $1$. 
$q$ (and, for asymmetric chains that are not invariant to reversal of the contour 
direction, its conjugate $q^\dagger$) 
is determined from a modified diffusion equation
\begin{equation}
\partial_s q\left(\vec{x},s; \left[\omega\right]\right) = \left[\nabla^2  - \omega\left(\vec{x}; s\right)\right]q\left(\vec{x},s;\left[\omega\right]\right).
\label{eq:diffusion}
\end{equation}
The $s$ dependence of the chemical potential field, 
$\omega$, is a feature of copolymers that arises from different species segments responding to different fields.
The most common initial condition for Eq.~\ref{eq:diffusion} is $q\left(\vec{x},s=0\right)=1$; $q^\dagger\left(\vec{x},s=1\right)=1$.

With the expression for $Q$ formalized, we turn our attention to the functional integrals
over fields in the canonical partition function, Eq.~\ref{eq:canpartfn}.
For many systems that are dense and far from phase transitions, the functional integrals
are dominated by saddle-point field configurations
\begin{equation}
\mathcal{Z}_C\approx \mathcal{Z}_0 \exp\left(-H\left[\left\{\omega^\star\right\}\right]\right).
\end{equation}
The self-consistent-field (SCFT) approach then reduces to finding the stationary, saddle-point field 
configurations, $\omega^\star$, for which the negative of the Hamiltonian functional, $-H$,
is real and large. 
A necessary condition for such a field configuration is for the Hamiltonian to be stationary in 
the sense that its functional derivative with respect to any of the fields
is zero:
\begin{equation}
\left.\frac{\delta H}{\delta \omega}\right|_{\omega^\star} = 0.
\end{equation}
For example, in the case of the incompressible diblock copolymer melt (Eq.~\ref{eq:ham_db}),
the mean-field equations are
\begin{eqnarray}
\left.\frac{\delta H}{\delta \omega_+}\right|_{\omega_+^\star} & =& 0 = i\frac{C}{\rho_0}\left(\rho_A\left(\vec{x}\right)+\rho_B\left(\vec{x}\right)-\rho_0\right)\nonumber,\\
\left.\frac{\delta H}{\delta \omega_-}\right|_{\omega_-^\star} & =& 0 = \frac{2C}{\chi_{AB}N}\omega_-^\star\left(\vec{x}\right) + \frac{C}{\rho_0}\left(\rho_B\left(\vec{x}\right)-\rho_A\left(\vec{x}\right)\right),\nonumber
\end{eqnarray}
where $\rho_0$ is the spatially averaged segment density, and $\rho_A$ and $\rho_B$ are spatially resolved segment densities of ``A'' and ``B''
components, which emerge from the functional derivatives of the normalized partition
function, $Q$, with respect to pressure and exchange fields:
\begin{eqnarray}
\rho_A\left(\vec{x}\right) & =& \frac{\rho_0}{Q}\int_0^{f_A} ds\, q\left(\vec{x},s\right)q^\dagger\left(\vec{x},1-s\right)\nonumber,\\
\rho_B\left(\vec{x}\right) & =& \frac{\rho_0}{Q}\int_{f_A}^1 ds\, q\left(\vec{x},s\right)q^\dagger\left(\vec{x},1-s\right),
\label{eq:dbdensity}
\end{eqnarray}
and $f_A$ is the fraction of the copolymer chain that is composed of ``A'' segments.
The first mean-field equation enforces local incompressibility of the melt, while the second encourages phase separation of the 
components --- which must be balanced against the loss of chain conformational and translational entropy.

Due to the complex nature of the field theory, care must be taken to ensure that the stationary
field configurations satisfying the mean-field equations 
have the correct saddle-point character, in that they correspond to 
local maxima for ``pressure-like'' fields that enter the Hamiltonian as $i\omega$, and to 
local minima for other fields.

In order to search for the mean-field configuration in an efficient and stable 
way, we use a dynamical relaxation scheme with appropriate Wick rotations applied to 
pressure-like fields.
To move beyond the mean-field approximation (SCFT) and affect full FTS simulations, 
we stochastically sample field configurations 
around the saddle-point using the complex Langevin (CL) scheme\cite{Klauder1983,Parisi1983}. This scheme ameliorates the sign
problem deriving from the complex ``probability distribution'' ($\exp\left(-H\right)$)
that leads to a critical loss of efficiency in Monte Carlo schemes.
The numerical aspects of these methods will be detailed in the following section.

\section{Algorithms}
With a model, and therefore a field-based Hamiltonian, defined, field-theoretic simulations
require a two-step iterative approach. 
The outer loop involves updating field configurations, either in search of a saddle-point 
configuration (dynamical relaxation), or through stochastic exploration of the functional 
integrals (complex Langevin). 
These types of simulation differ only in that the latter contains a stochastic driving term in the
equation of motion of the fields:
\begin{equation}
\frac{\partial \omega}{\partial t} = -\lambda\frac{\delta H}{\delta\omega} + \eta\left(\vec{x},t\right),
\label{eq:EOM}
\end{equation}
where it is understood that here the fields, $\omega$, are not Wick rotated, and 
where the final term is omitted for mean-field SCFT calculations.
The stochastic driving, $\eta$, is a real, Gaussian noise that is decorrelated in both space and
time, and is defined by its first and second moments, which are selected to reproduce the correct
time-averaged distribution using the fluctuation-dissipation theorem: $\left<\eta\left(\vec{x},t\right)\right>=0$
and 
$\left<\eta\left(\vec{x},t\right)\eta\left(\vec{x}^\prime,t^\prime\right)\right>=2\lambda\delta\left(\vec{x}-\vec{x}^\prime\right)\delta\left(t-t^\prime\right)$.

In order to solve for the time evolution of the chemical potential fields with discrete time stepping of 
the equation of motion (Eq.~\ref{eq:EOM}), we employ recently developed stable and accurate integration schemes.
For SCFT calculations, the most important numerical consideration is a wide stability window for large time steps.
Stable numerical schemes admit large time steps and therefore the rapid advancement of the fields to a saddle-point
configuration, while details of the trajectory taken are less important.
For this task, we have found the SIS method introduced by Ceniceros and Fredrickson\cite{Ceniceros2004} to be very efficient.
In this approach, the linearized response of the density fields appearing in $\delta H/\delta \omega$ are treated 
semi-implicitly to confer large gains in stability over standard forward-Euler propagation.
CL simulations, on the other hand, require \emph{accurate} discretization schemes for time stepping in order to 
compute reliable time-averaged equilibrium quantities. 
Achieving accurate time propagation for simulations in which fluctuations are large ($\eta \sim \delta H/\delta \omega$)
is not trivial.
For this task, the recently introduced exponential time differencing (ETD) algorithm\cite{Villet2010, VilletUnpublished}, which incorporates 
the linearized force into an integrating factor, has proven to be accurate and efficient.
Still, with even the most efficient algorithms, the emphasis on time-step accuracy, and the requirement to collect
stochastic averages of operators, necessitates that CL calculations in the strong fluctuation limit 
usually require one to two orders of magnitude more runtime than SCFT calculations, making fluctuating field-theoretic
simulations an ideal candidate for acceleration with GPUs.

To further formalize the discrete field-update algorithms, we 
discuss in detail only the example of a first-order semi-implicit scheme\cite{Lennon2008SIAM}, which reduces to SIS in the limit of zero noise.
The final discretized time-stepping expressions in reciprocal space are:
\begin{eqnarray}
\hat{\omega}^{t+\Delta t,\vec{k}}  & = & \hat{\omega}^{t,\vec{k}} - \lambda\Delta t\left[\frac{\delta H\left[\hat{\omega}^t\right]}{\delta\hat{\omega}^{t,\vec{k}}} 
+ L\left(\frac{\delta H\left[\left\{\hat{\omega}^{t+\Delta t}\right\}\right]}{\delta \hat{\omega}^{t+\Delta t,\vec{k}}}\right)\right]\nonumber\\
&+& \lambda\Delta t  L\left(\frac{\delta H\left[\left\{\hat{\omega}^{t}\right\}\right]}{\delta \hat{\omega}^{t,\vec{k}}}\right) + \mathcal{F}\left(\bar{\eta}^{\vec{x},t}\right),
  \label{eq:SIS}
\end{eqnarray}
where $L$ implies taking the linear part of the force which, 
in the weak inhomogeneity limit, is typically of the form $\Delta \rho \approx g\star \omega$, where
$g$ is a Debye function that depends on the polymer architecture.
$\mathcal{F}$ is the discrete Fourier transform, and $\bar{\eta}$ is the \emph{discretized} Gaussian noise, which has variance
$\left<\bar{\eta}^{\vec{x},t}\bar{\eta}^{\vec{x}^\prime,t^\prime}\right> = 2\lambda\Delta t\delta_{t,t^\prime}\delta_{\vec{x},\vec{x}^\prime}/\Delta V$.
Treating the linear part of the force semi-implicitly leads to an effective wavevector-dependent time step:
\begin{equation}
  \hat{\omega}^{\vec{k},t+\Delta t} = \hat{\omega}^{\vec{k},t} - \frac{\lambda\Delta t}{1+\lambda\Delta t \hat{g}^{\vec{k}}}\frac{\delta H\left[\hat{\omega}^{t}\right]}{\delta \hat{\omega}^{\vec{k},t}} + \hat{\bar{\eta}}^{\vec{k},t}.
\end{equation}
Any explicitly linear terms in the force expression are handled fully implicitly in the time-stepping algorithm, which
further modifies the latter expression. 
We note here that field updates are entirely local in reciprocal space, even accounting for the linearization terms,
so that the time propagation is insensitive to the choice of simulation boundary conditions for conditions compatible with
plane wave and Fourier bases.  
We also note that the field update equations are 
order-$M$ (linear scaling with grid dimension, $M$), local, and trivially parallelizable over the Fourier spatial modes.

Within the inner loop of a field-theoretic simulation, the modified diffusion equation (Eq.~\ref{eq:diffusion})
for the chain propagators, $q$, must be solved for the static field configurations so that the 
normalized partition functions and spatially resolved segment densities are available.
For this task, we have found pseudo-spectral\cite{Rasmussen2002,Tzeremes2002} methods based on operator splitting 
to be the most efficient schemes. 
Specifically, we use an extension to the original pseudo-spectral approach that cancels the lowest order error using
Richardson extrapolation, leading to a scheme that is globally fourth-order accurate in the step size of the
discretized contour variable, $\Delta s$\cite{Ranjan/Qin/Morse:2003}.
The discrete solution for a single contour step may be written
\begin{widetext}
\begin{eqnarray}
q\left(\vec{x},s+\Delta s; \left[\omega\right]\right) & = & -\frac{1}{3}\left[e^{\left(-\omega\left(\vec{x}\right)\Delta s/2\right)}  e^{\left(\nabla^2 \Delta s\right)}  e^{\left(-\omega\left(\vec{x}\right)\Delta s/2\right)} \right. \nonumber \\
    & - & \left. 4 e^{\left(-\omega\left(\vec{x}\right)\Delta s/4\right)}  e^{\left(\nabla^2 \Delta s/2\right)}  e^{\left(-\omega\left(\vec{x}\right)\Delta s/2\right)} e^{\left(\nabla^2\Delta s/2\right)} e^{\left(-\omega\left(\vec{x}\right)\Delta s/4\right)}\right]  q\left(\vec{x},s;\left[\omega\right]\right) + \mathcal{O}\left(\Delta s^5\right),
  \label{eq:RSOS}
\end{eqnarray}
\end{widetext}
where the application of the diffusion operator, $\exp\left(\nabla^2\Delta s/2\right)$, occurs in Fourier space by forward and reverse fast Fourier transform (FFT) operations on the propagator.
Since the chemical potential field acts locally in real space, and the diffusion operator acts locally in Fourier space, both operators are 
diagonal in the relevant representation, which makes evaluation of 
the exponentiated operators a numerically simple task.
As we will show elsewhere\cite{AudusUnpublished2011}, this scheme offers the best performance of our available methods, 
as measured by the wall-clock time taken to reduce the 
contour discretization error to an arbitrary pre-specified level\cite{StasiakMatsen2011}.
Since the modified diffusion equation is semi-local, boundary conditions must be specified. 
Pseudo-spectral schemes can be easily developed for Dirichlet, Neumann and periodic boundary conditions 
by using discrete sine, cosine or Fourier transforms respectively. 
A Chebyshev-based pseudo-spectral approach is available for more general types of boundary conditions\cite{Hur2012}.

\section{Structure of the Field-theoretic Simulation Code}
Our polymer field-theory code was written in C++ with strict object-oriented design principles and type templating.
Data is private and strictly encapsulated within each class, while an exposed public interface facilitates
manipulation. 
This strategy allows a strong separation between the internal representation of any
data structures and the operations that can be conducted on that data, which guarantees that the code for running 
on GPUs, with its inaccessible memory spaces, or another parallel architecture, requires relatively few changes.

Inheritance is employed throughout the code to expose a common interface for objects of classes
derived from a common parent, e.g., the chain propagators for homopolymers, diblock copolymers, asymmetric multiblock polymers
and other architectures are used identically after their creation. 
This feature provides a flexible framework for tackling a large body of field-theory problems.

Our class hierarchy can be broadly divided into low-level and high-level classes. 
Within the low-level partition are classes that represent device-dependent functionality, 
such as an \verb!FFTvector! container for manipulating all functions of a single spatial variable
that are subject to periodic boundary conditions.
Through inheritance, these classes are targeted to run on a variety of platforms with highly optimized 
code for GPUs, and for CPUs exploiting OpenMP (shared memory) and/or MPI (distributed memory) parallelism. 
The low-level classes also include overloaded arithmetic operators to give a simple programming model for manipulating
fundamental objects within the high-level classes. 
However, for optimal efficiency, we avoid binary operators (\verb$*$, \verb$+$, \ldots), which necessitate creation of temporary
objects. 
Such a requirement imposes a high operational overhead when temporary objects can be tens or hundreds of megabytes in size, 
and we therefore take care to restructure all steps of our 
algorithms using only compound assignment operations (\verb$*=$, \verb$+=$, \ldots) and minimal object copies.

The high-level partition of our code includes classes that define the components of various simulation types, including
solving for the chain propagator, computing the normalized partition function and segment density, and 
stepping the discretized equation of motion with various ``plug-in'' field updater algorithms.
Since the code implementing rules of arithmetic are contained entirely within the low-level partition, the 
code in the high-level partition is ignorant of the details of the simulation platform. 
Simulations running on a GPU, serial CPU, or parallel CPUs have identical high-level code, as do simulations employing
single- or double-precision floating-point arithmetic.

Developing code within the high-level partition can lead to the introduction of inefficient practices.
We avoid this problem with careful code design and profiling, ensuring that our code is FFT dominated on all platforms, and
providing for a fair comparison between the runtimes for simulations running on GPU or CPU.
For example, to achieve maximum throughput in solving the discretized diffusion equation, all exponentiated 
field operators used in the pseudo-spectral operator-splitting scheme (Eq.~\ref{eq:RSOS}) are pre-computed when the propagator 
initial conditions are set. 
Furthermore, specifically for the diffusion operator, pre-computation may occur just once per simulation provided the 
set of plane waves does not change (i.e., in the absence of variable-cell methods, which we do not consider in the present work). 
Similarly, the linear and linearized force coefficients for the field time stepping (Eq.~\ref{eq:SIS}) are computed only once
per simulation.

\subsection{Serial and Parallel CPU Implementation}
\label{sec:cpu}
Our simulation code is parallelised over plane waves or real-space collocation grid points.
For implementing the FTS framework, three distinct parallel patterns appear: Fourier transformations, direct vector operations,
and reduction operations.

Fourier transformations, which are required to solve the modified
diffusion equation pseudo-spectrally, and to make SIS and ETD field updates
local, are handled in CPU-targetted code by the freely available \verb!FFTW! library\cite{FFTW}, 
which has serial and both SIMD and MIMD parallel execution capabilities.
We follow established conventions for data distribution with \verb!FFTW!.
We compiled the \verb!FFTW! library with SSE extensions, and thoroughly tested the performance of the
library with respect to various compiler optimizations for each target platform.
For simulations running in parallel, three different strategies are available: shared-memory parallelism with 
OpenMP\cite{Note1}, 
distributed-memory parallelism with explicit message passing using MPI, and a combination of the two. 
With a dual-parallel execution model, it is important to be sure that CPU cores do not become oversubscribed by the launch of more threads than there
are available cores.

Direct vector operations, used for example to apply exponentiated operators in the pseudo-spectral scheme, or to 
time-step fields once ``forces'' are fully determined, are trivially parallelisable, requiring no thread or process cooperation.

Reduction operations are used throughout our code to perform spatial integrations (e.g., in evaluating the normalized partition
function, Eq.~\ref{eq:Q}, or operators such as the energy functional, Eqns.~\ref{eq:ham_edwards} and \ref{eq:ham_db}), 
or for evaluating any type of norm of a field or function. 
We implement reduction operations within MPI using \verb!MPI_Allreduce! operations, and within OpenMP using private variables 
for thread-level block reductions, followed by atomic updates of a shared result variable.

We use version 12 of the Intel C++ compiler wrapped with OpenMPI for compiling our CPU-target code.

\subsection{General-Purpose Computing on Graphics Processing Units}
GPU architectures are optimized for high throughput computing rather than peak
single-thread performance.
They consist of a collection of symmetric multiprocessing units, each of which has 
a number of thread processors with distinct arithmetic logic units.
For example, NVIDIA T20 series has $14$ multiprocessors with $32$ cores per processor, for a total of $448$ cores.
GPU cores are relatively lightweight with simple instruction sets, and the hardware lacks many performance aids present
in modern CPUs, such as multiple pipelines and large on-chip caches.
However, thread support is implemented in hardware, so that thread creation, destruction, switching and synchronization 
are low-overhead operations.

In principle, code can be written to target simultaneous GPU and CPU execution with load sharing. 
However, in most cases this would inevitably require explicit data transfer through the PCI-express bus, which
imposes a high penalty for simulation codes handling large data sets. 
For field theoretic simulations, we have yet to find any performance benefit in such load sharing.

While modern GPUs have a large store of shared device-level DRAM ($3$--$6$\,GB on NVIDIA Tesla T20 series), memory accesses are 
relatively slow with high latency. 
This dictates a powerful strategy for achieving high performance: the GPU should be overloaded with threads. 
Threads that are delayed due to memory latency are switched out by a scheduler in favor of those that are ready to run. 
The switching overhead is negligible, so in this way the memory latency can be ``hidden'' by overloading the GPU.

Threads are launched in groups, called blocks, of size determined by code implementation.
Efficient thread communication occurs only \emph{within} blocks by individual threads 
loading and storing of data in a small reserve of on-chip shared memory.
The shared memory is usually very limited, typically less than $64$\,KB, but accesses are very fast. 
Hence, in addition to being used for thread cooperation, the shared memory space can be used as a temporary store of 
data fetched from global device memory, to avoid subsequent repeat reads, and an important aspect of code optimization is
the careful use of this limited resource.
The blocking of threads imposes an important constraint for translating algorithms into GPU code: the
order of block execution is undefined due to dynamic thread scheduling, so that cooperation between threads in different
blocks is difficult and inefficient.
For best performance and robust code, the computation should not depend on the order in which blocks are executed.

At any given snapshot in time, threads will be running in multiples of the warp size ($32$), regardless of the block size. 
Threads that are executing the same instructions concurrently will benefit with improved performance, as 
will threads that read neighboring memory addresses in device memory. 
Hence, minimizing thread ``divergence'', in terms of executed instructions and memory access patterns, 
yields optimal computational throughput.

The general programming frameworks for GPUs that have recently been developed, CUDA and OpenCL, provide a small set of 
extensions to existing programming languages. 
Host (CPU) code is compiled and executed as before, but extensions are included for orchestrating the 
launch of GPU ``kernels'' from the host code.
With the release of these frameworks, GPU programming is significantly simplified, and the development of efficient
code can be achieved by noting the small set of considerations listed above.

\subsection{CUDA implementation of Field-Theoretic Simulations}
\label{sec:cuda}
Given the design and optimization constraints outlined in the previous section, two crucial design principles emerge.
Efficient GPU use requires many threads to hide memory latency, each of which may be created and destroyed for the sole purpose of executing a single 
arithmetic operation on a single element of data.
Furthermore, data transfers between host system memory and GPU device memory should be infrequent.
For these reasons, our field-theoretic code is fundamentally designed to permanently store all quantities on the GPU device memory.
Transfers of data to host system memory only occur for the infrequent output of instantaneous operator values and density-field snapshots.
In addition, where possible, each thread during the execution of GPU kernels handles 
only a single plane-wave coefficient or real-space collocation sample of a function.

For the parallel patterns discussed in Sec.~\ref{sec:cpu}, we have written efficient CUDA kernels for parallelising the 
operations over individual spatial samples.
The exception is the Fourier transformations, which are handled by an efficient library, \verb~cuFFT~, 
included with the CUDA software development kit.
Thread computations within this library are by far the dominant part of our simulation runtime.

Complex Langevin calculations require generation of the Gaussian noise 
data for every field update (see Eq.~\ref{eq:SIS}). 
The normally distributed noise can be generated with thread-wise parallelism
on the GPU using the CUDA ``CURAND'' library. 
The net result on runtimes is an overall $\sim 10\%$ increase in 
performance for CL simulations, compared with running the simulation on the GPU
but generating the Gaussian noise in serial on the CPU.

Considering the vastly increased, but still limited, 
global memory size of $3$--$6$\,GB of DRAM available on modern GPUs, a brief analysis 
of the memory costs of an FTS calculation is in order.
Taking the example of a diblock copolymer melt that contains only a single distinct forward-chain propagator, we 
assume $100$ contour steps ($\Delta s$) are required to achieve well-converged results for Eq.~\ref{eq:RSOS}.
\cite{Note2}
%
Assuming all spatially dependent functions are resolved with $128$ plane waves in each of three spatial
dimensions, corresponding to a large calculation, a single function of space represented with $128$-bit 
double-precision complex numbers requires $32$\,MB of storage.
Thus, the chain propagator requires $\sim3$\,GB of data.
Upon introduction of the conjugate propagator, which is the same size, 
we already exceed the storage available on the largest GPUs.
Each field and 
exponentiated operator, of which there are relatively few, also requires $32$\,MB.
To save memory, we do not store $q$ and $q^\dagger$ separately, but rather the quantity
$q\left(s\right)q^\dagger\left(1-s\right)$. 
This quantity is computed and stored on-the-fly during the diffusion computation, so 
that the individual forward and conjugate propagators are never required to be
stored individually.
Provided the propagators are only used to calculate the partition function and the density field, 
this quantity is all that is required, which allows for a $50$\% reduction in memory cost. 
(Note that other properties, such as the osmotic pressure or stress tensor, are more complicated functionals of the chain propagators, and
 this memory-saving measure cannot be employed when such quantities are required). 
With this choice, almost all present simulations of interest can fit comfortably into $6$\,GB of global storage. 
However, for the rare case that the GPU device memory is exhausted, we have implemented the option to move 
the quantity $q\left(s\right)q^\dagger\left(1-s\right)$ into host system memory as it is being computed. 
This saves $\gtrsim 95$\% of the memory cost of a typical calculation, at the expense of some performance loss.

\section{Results and Benchmarking}
\label{sec:results}
For our performance testing, we compare code runtimes on late-model CPUs and GPUs deployed in the \verb!knot! cluster at
the California Nanosystems Institute at the University of California, Santa Barbara.
The CPUs used in our tests were Intel\textsuperscript{\textregistered}
six-core Xeon\textsuperscript{\texttrademark} E5650 processors clocked at
$2.66$\,GHz. 
Each node of the cluster has two such six-core CPUs installed on the motherboard, for a total 
shared-memory parallel capacity of 12 cores. 
115 nodes are connected in parallel using fast Infiniband
interconnects (MT26438 from Mellanox Technologies) for a total 
parallel capacity of 1380 cores.

For the tests conducted purely with MPI-level parallelism, MPI processes can 
execute concurrently either on cores within the same CPU/node, or on CPU cores located in different nodes. 
In either case the parallel cooperation proceeds through explicit message passing, regardless of
whether the cooperating cores have access to a common shared memory pool. 
For MPI computations involving more than 12 processes, the simulation inevitably must 
execute on multiple nodes with message passing through the Infiniband interconnects, but with 
few processes we find that it is advantageous to run on a single node.
This indicates that competition between the MPI processes for memory bandwidth provides less of a
overhead than message passing between nodes.
In contrast, with OpenMP parallelism execution threads may \emph{only} cooperate through
a shared memory pool (typically $24$ gigabytes on \verb!knot!) and such simulations are therefore limited to $12$ cores 
the \verb!knot! cluster.
Finally, for mixed \verb!OpenMP+MPI! parallelism, we exploit shared-memory cooperation within a node
and explicit message passing for cooperation between nodes. 
In this case, each node handles two MPI process (one for each multi-core CPU), which
are responsible for orchestrating branching and merging of multiple OpenMP threads (up to the
maximum of $12$ per node).

The GPUs used in our tests were NVIDIA Tesla C2050 boards based on the
Fermi architecture with 3GB global DRAM. 
We turn off the error-correcting code (ECC) feature on the global memory storage of the GPU, 
leading to faster memory transactions.
We observe an overall 10\% performance boost with ECC turned off.
Unless otherwise stated, all of our timings are for simulations using double-precision
arithmetic.

\begin{figure*}[H!tp]
\begin{center}
\includegraphics[width=0.85\textwidth,keepaspectratio=true]{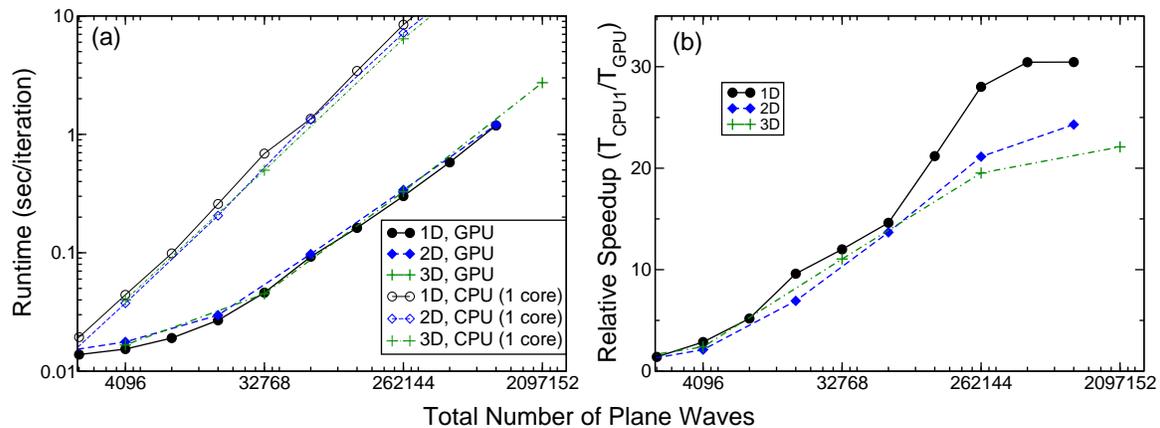}
\caption{(Color online) Benchmarking data for SCFT calculation of LAM phase in 
  1D, 2D, and 3D. In all simulations $f_A = 0.5$ and $\chi_{AB} N = 14.0$ in a cubic cell with side
  length $10.89$\,$R_g$, and fields were initially seeded with decorrelated random noise.
(a) Absolute timings in seconds per field update (propagator calculation + field stepping).
The CPU code is executed in serial on a single CPU core with on-core SSE vectorizations enabled.
Our GPU code can perform one full field update in $\sim 1$\,sec.~for even very
large basis sets corresponding to $128^3$ plane waves.
(b) Relative speedup of GPU run over a single CPU core. All timings are averaged over three runs, 
  with each simulation $1000$ field updates in length.
Lines are a guide to the eye, and are not indicative of runtime or scaling for
non-power-two simulation grids.
See body text for hardware details.
}
\label{fig:LAMtimingsSCFT}
\end{center}
\end{figure*}

Fig.~\ref{fig:LAMtimingsSCFT} shows timing data for self-consistent field
calculations of the lamellar (LAM) phase of an incompressible melt of diblock
copolymer.
For this case, the calculations were run either on a 
\emph{single CPU core} (serial execution, albeit with some on-core vectorization 
in the form of SSE instructions) or a single GPU (threaded parallel execution).
Panel (a) shows absolute timing data with respect to the total number of 
plane waves used to represent the field for 1D, 2D and 3D simulations.
The GPU timings are clearly significantly shorter than simulations running
on a single CPU core for all but the smallest of calculations, which suffer
performance penalties due to not being able to overload the GPU with
sufficient threads to hide memory latency. 
The GPU simulations require $\lesssim 1$\,sec.~for a single 
field update, even for more than one million plane waves.
We also find that the GPU and CPU performance does not depend significantly
on the dimensionality of space.

Fig.~\ref{fig:LAMtimingsSCFT}(b) shows the relative speedup of the GPU
code over the serially executed CPU code, defined as $T_\mathrm{CPU1}/T_\mathrm{GPU}$.
The performance gap widens with increasing simulation size, and the peak 
speedup we observe is greater than $30\times$.
Our largest calculation was a 3D simulation with $128$ plane waves per dimension,
for a total of $2,097,152$ spatial degrees of freedom. 
Larger simulations are possible, but they become increasingly challenging from 
the perspective of obtaining 
accurate timing statistics for single-core serial CPU references due to the long runtimes involved.

\begin{figure}[H!tp]
\begin{center}
\includegraphics[width=0.95\columnwidth,keepaspectratio=true]{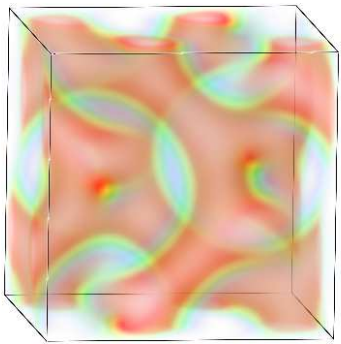}
\caption{(Color online) Volumetric plot of the minority-component density of a diblock copolymer melt in the 
in the gyroid phase. The chosen Flory interaction parameter is, $\chi_{AB}N=14.4$, and minority block fraction
is $f_A = 0.4$. The simulation domain is of size $8.82$\,Rg in each dimension. Timings for this simulation
are shown in Fig.~\ref{fig:GYRtimingsSCFT}.}
\label{fig:gyrmorph}
\end{center}
\end{figure}

\begin{figure*}[H!tp]
\begin{center}
\includegraphics[width=0.85\textwidth,keepaspectratio=true]{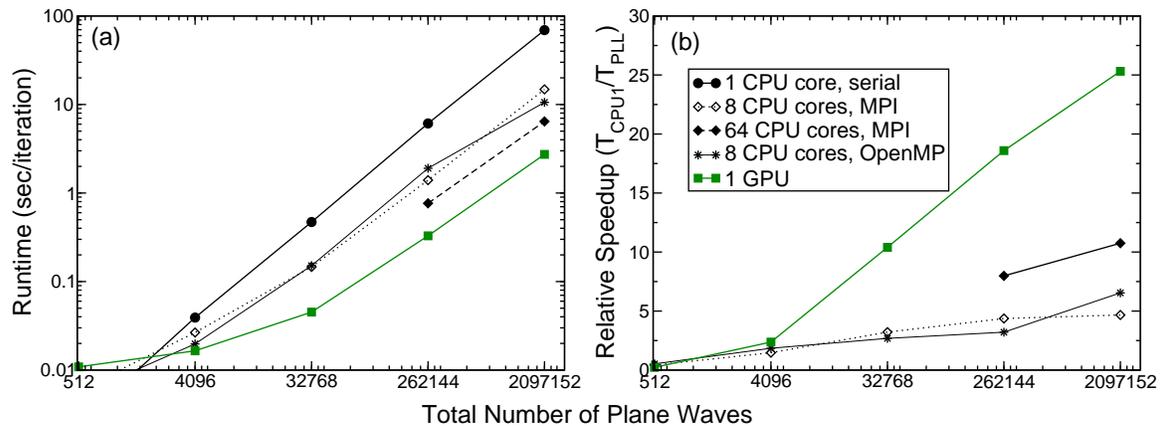}
\caption{(Color online) 3D SCFT simulation of the cubic gyroid phase of 
  an incompressible diblock copolymer melt. In all simulations, $f_A = 0.4$, 
  $\chi_{AB}N = 14.4$ and the simulation cell is a cube with side $8.82$\,Rg$^{-1}$.
  (a) Absolute timings in seconds per field iteration (diffusion + field stepping).
  (b) Relative speedup of GPU and parallel CPU implementations over a single CPU.
  $T_{PLL}$ refers to the timing on any of the parallel platforms, including GPU.
All timings are averaged over three runs, with each simulation $1000$ field updates in length.
Lines are a guide to the eye, and are not indicative of runtime or scaling for
non-power-two simulation grids.
See body text for hardware details.
}
\label{fig:GYRtimingsSCFT}
\end{center}
\end{figure*}

To demonstrate that the favorable GPU scaling we observed in the previous tests does 
not derive from a fortuitous benefit of phases such as LAM with translational invariance
in two of the spatial dimensions, we reran benchmarking tests on the 
gyroid phase of the incompressible diblock copolymer melt\cite{Cochran2006}. 
This phase, shown in Fig.~\ref{fig:gyrmorph}, has no unrestricted translational invariance and 
cannot be represented in fewer than three spatial dimensions.
The results shown in Fig.~\ref{fig:GYRtimingsSCFT} expose a similar $>20\times$
peak speedup over a single CPU core as found for simulations in the LAM phase.
In this case, we also compare to parallel CPU simulations conducted both
with shared-memory (OpenMP) parallelism, and with distributed-memory
(MPI) parallelism. 
We note that when all MPI processes are launched on a single shared-memory node
when possible, the performance of our MPI implementation is comparable to that
of our OpenMP implementation, indicating a low overhead of explicit message
passing in this limit.
Presumably a result of the high communications overhead of the 
transposing steps taken in distributed-memory implementations of 
multi-dimensional FFTs, we find an MPI performance that saturates at
$\sim10\times$ speedup even for 64 CPUs.
The result is that a single GPU outperforms even one hundred CPUs cooperating
via MPI (not shown) for our FFT-dominated method.

\begin{figure}[H!tp]
\begin{center}
\includegraphics[width=0.85\columnwidth,keepaspectratio=true]{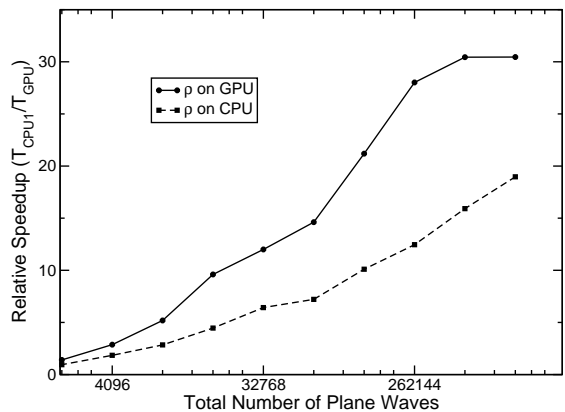}
\caption{GPU calculation speedup over a single CPU core for the 
1D LAM phase shown in Fig.~\ref{fig:LAMtimingsSCFT}. Comparison of 
full simulation conducted on the GPU to a calculation in which the density
is calculated using the CPU. The latter option significantly saves GPU device
memory with some performance loss due to data transfer overhead and loss of parallelism.}
\label{fig:rhoCPU}
\end{center}
\end{figure}

As discussed in Sec.~\ref{sec:cuda}, for the rare simulations that exceed the GPU 
global shared memory capacity we have 
implemented the option to stage the quantity $q\left(s\right)q^\dagger\left(1-s\right)$ on-the-fly
into host memory following each contour step in solving the discretized modified 
diffusion equation, Eq.~\ref{eq:RSOS}.
Subsequently, the density fields are calculated in serial on a single CPU core through 
integrations over the contour variable (e.g., Eq.~\ref{eq:dbdensity}).
With this option, the memory allocation on the GPU is limited to a small number of 
individual fields and operators, while the much larger $s$-dependent propagators are never
stored locally.
While this option typically saves more than $95$\% of the memory, depending on 
the underlying model, it also incurs some performance penalty both due to loss of parallelism on
performing the contour integrals to evaluate the density, and from the cost of 
transferring gigabytes of data to host memory during the diffusion contour stepping.
Fig.~\ref{fig:rhoCPU} shows the performance loss from this memory-saving option,
which amounts to approximately $50$\% loss of the peak acceleration obtained from
running entirely on a GPU.

\begin{figure}[H!tp]
\begin{center}
\includegraphics[width=0.85\columnwidth,keepaspectratio=true]{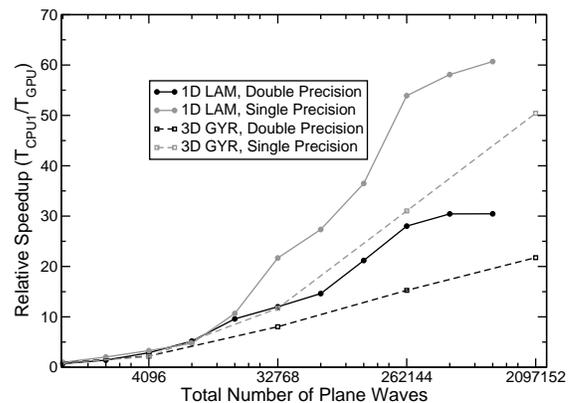}
\caption{Relative speedup of GPU calculations conduction with single-precision floating-point arithmetic
  (32 bits per float) proceed approximately two times faster than those using double precision. The reference
  timings are single-core serial CPU calculations using \emph{double-precision arithmetic}. 
  Numerical and model details of the LAM and GYR phase calculations shown here are the same of those
  used in Figures \ref{fig:LAMtimingsSCFT} and \ref{fig:GYRtimingsSCFT} respectively.}
\label{fig:sprec}
\end{center}
\end{figure}

Due to the reliance of our simulation code on type templating, switching arithmetic from double-precision
(64-bit floating point numbers) to single-precision (32-bit floats) is a simple task.
Before the launch of the Fermi architecture, NVIDIA GPUs did not have hardware support for double-precision arithmetic, and even 
the latest hardware suffers a FLOP throughput reduction of a factor of two when performing double-precision arithmetic.
Hence, the ability to run calculations in single precision is desirable both to support older hardware, and to 
further improve performance.
While single-precision arithmetic carries an associated increase in numerical error, which may accumulate during the solution of our 
non-linear differential equations, the advanced stable algorithms that we use sufficiently suppresses serious error 
accumulation and consequent destabilization.
The utility of single-precision arithmetic in SCFT simulations is obvious for rapidly evolving to the saddle point solution of the
mean-field equations. 
Further refinement with double-precision processing can be performed if desired once the field iteration has converged in single precision.
The runtime speedup for LAM and GYR phases using single-precision arithmetic on a Tesla C2050 GPU are shown in Fig.~\ref{fig:sprec}, with
the reference runtime being the double-precision timings for serial CPU execution. 
We note that the CPU-only execution is not significantly accelerated when running with single-precision arithmetic, despite the use of SSE3
instructions.

\begin{figure}[H!tp]
\begin{center}
\includegraphics[width=0.90\columnwidth,keepaspectratio=true]{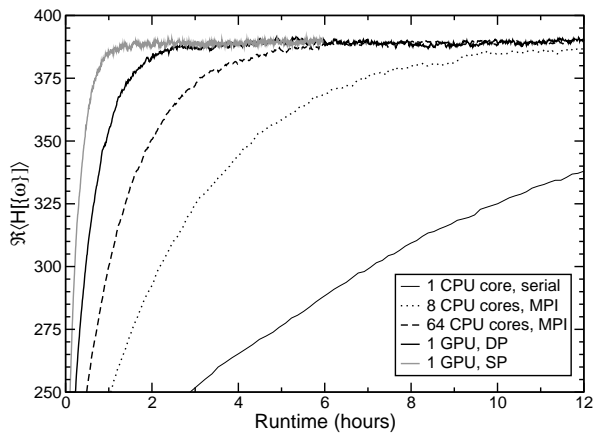}
\caption{CL dynamical trajectory of $\Re \left<H\right>$ plotted against wall-clock runtime for the
simulation of a single cubic unit cell of the gyroid phase of an incompressible diblock copolymer melt. 
The simulation parameters are the same as for the mean-field calculations of the same phase shown in 
Fig.~\ref{fig:GYRtimingsSCFT}. For this example, we choose to represent fields with
a grid of $64^3$ plane waves. The relative strength of fluctuations is determined by $C$, which we
set to $50$ for this simulation. Operator values were saved after every $100$ field updates. Single-precision (SP)
GPU calculations progress approximately two times faster than double precision (DP) equivalents.}
\label{fig:gyroidCL}
\end{center}
\end{figure}

Fig.~\ref{fig:gyroidCL} shows a comparison of serial, MPI and GPU runs
of a complex Langevin simulation of the cubic gyroid phase for relatively strong
field fluctuations (determined\cite{Ganesan2001} for the model of interest by the chain-density 
parameter $C=50$). 
For this case, we plot the dynamical trace of the real part of the Hamiltonian operator.
Once equilibration of the CL such trajectories has been achieved, 
time averages correspond to weighted averages over field configurations, and therefore
to expectation values of the averaged operator.
Note that while the average of the Hamiltonian operator, $\left<H\right>$, 
is loosely linked to the Helmholtz free energy\cite{Lennon2008}, these quantities are
not equal because $\left<H\right>$ does not include all entropic contributions from 
field fluctuations.
However, this quantity remains a useful measure of equilibration of the CL trajectory.
Achieving fast equilibration of CL trajectories for large 3D simulations is a
major challenge, and one that is only partially resolved by improved algorithms 
such as ETD.
Fig.~\ref{fig:gyroidCL} demonstrates that the $20$--$30\times$ performance gain provided
by running on GPUs has a significant impact on the our ability to quickly move through
the warmup stage of our simulations, with warmup achieved more than two times faster
than $64$ MPI processes running on $64$ CPU cores.
Single-precision arithmetic can be used to further accelerate the warmup process by an additional factor of two.
In the example shown, the operator averages in single- and double-precision match to with statistical
error, though even if this were not the case, the equilibriated simulation could be continued accurately
by switching to double precision after the warmup stage.
Note that the simulation cell used in this example is the smallest possible for the cubic gyroid
phase, and fully capturing the effects of spatially decorrelated fluctuations may require larger simulation cells, 
which can require days to reach equilibrium with traditional MPI codes.

\begin{figure}[H!tp]
\begin{center}
\includegraphics[width=0.90\columnwidth,keepaspectratio=true]{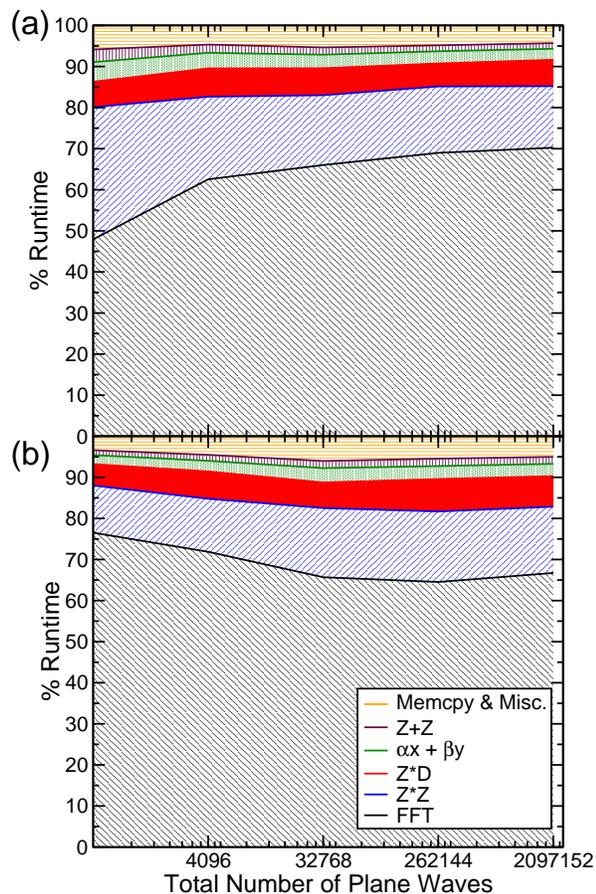}
\caption{(Color Online) Flat profiles showing the breakdown of serial single-core CPU (panel `a') and GPU (panel `b') runtimes for the different 
activities performed during an SCFT simulation of the gyroid phase, for which absolute timings were 
reported Fig.~\ref{fig:GYRtimingsSCFT}. The activities are: Fast Fourier transformations (``FFT'');
complex-complex vector multiplies, used in field operator application (``$Z*Z$''); complex-real vector multiplies, used in
application of the Laplacian operator in reciprocal space (``$Z*D$''); the sum of 
two scaled vectors, used in Richardson extrapolation of the operator-split diffusion equation (``$\alpha x + \beta y$''); and 
summation of two complex vectors, used in contour integration to obtain species density fields and in evolving the 
field equation of motion (``$Z+Z$''). The remaining runtime is consumed by memory copies and miscellaneous CPU-code execution.}
\label{fig:GPUprof}
\end{center}
\end{figure}

Finally, we analyze the flat profile of serial single-core CPU and GPU runtimes shown in Fig.~\ref{fig:GPUprof}(a) and (b)
  respectively for the gyroid calculations presented in Fig.~\ref{fig:GYRtimingsSCFT}.
The profile data demonstrates that the GPU runtime is dominated by fast Fourier transformation
operations, with approximately $70\%$ of runtime consumed by this single activity. 
The remaining largest consumers of runtime are element-by-element products of field data for
complex-complex multiplies ($Z*Z$) or complex-real multiplies ($Z*D$). 
These three operations collectively account for $>90\%$ of runtime on all simulation sizes.
The latter two, involving multiplies within CUDA kernels, are strongly bandwidth limited 
because multiple read/write operations are required for each arithmetic operation. 
We note that the design strategy of restricting all data to the GPU device memory, and not utilizing the
host storage, results in less than $5\%$ of runtime being spent on CPU code and memory copies.
We find similar profile data for our CPU-executed code, with the exception that FFT costs reduce to only $\sim 50\%$ for
the smallest of system sizes.

\section{Conclusions}
We have demonstrated that field-theoretic simulations running on a recent model 
graphics
processing unit can progress up to sixty times faster than those running in serial on 
a single core of a recent model CPU. 
This enhanced performance was observed in simulations conducted in the 
mean-field (SCFT) limiting case, as well as in complex Langevin simulations 
of the fully fluctuating field theory.
Due to the intensive communications overhead associated with fast Fourier transforms,
the same code running on CPUs in parallel, either with shared or distributed
memory strategies, saturates with approximately ten times speedup over a serial
calculation.
Hence, a single GPU currently provides the highest throughput capacity for difficult 
field-theory problems, even when compared to the largest CPU-based clusters.
The performance gains that we observe make fully fluctuating, large-scale 3D
simulations of complex phases much more tractable.
Since GPU technology is evolving rapidly, and compute throughput is increasing
significantly with each generation, we anticipate that future hardware developments or enhancements
to the CUDA numerical libraries will 
provide further significant ``drop-in'' gains in performance with little or no extra code
development.

While our peak GPU performance is very favorable, we note that small calculations do
not benefit from running on a GPU. 
In such cases, it is not possible to provide sufficient threads to hide memory 
latency.
We find the break-even point to be somewhere between $512$ and $2048$ total plane waves,
which is a smaller basis size than is required for almost any problem of interest
involving large simulation cells.

Strategies for further improvements in performance could include sharing
the computation load between CPU and GPU, identifying strategies for exploitation of
asynchronous GPU kernels, and adding additional levels of parallelism so that calculations
can run simultaneously on multiple GPUs. 
However, due to slow memory transfers and high communications overhead, it is unclear
whether these strategies would provide significant further gains in performance.

\begin{acknowledgments}
KTD was supported by the National Science Foundation SOLAR program (CHE-1035292) and
GHF received support from NSF award DMR-0904499.
This work was partially supported by the MRSEC Program of the National Science Foundation under Award No. DMR 1121053
and made use of the 
California Nanosystems Institute Computing Facility with resources provided 
by NSF award CNS-0960316.
We thank Paul Weakliem of the CNSI Computing Facility for hardware support 
and fruitful discussions.
\end{acknowledgments}

\bibliographystyle{apsrev}

\end{document}